# Image-to-Image Translation Based on Deep Generative Modeling for Radiotherapy Synthetic Dataset Creation

Olga Glazunova[1], Cecile J.A. Wolfs[2] and Frank Verhaegen[2]

[1] Health as Digital Transformation, Faculty of Health, Medicine and Life Sciences, Maastricht University, Maastricht, the Netherlands
[2] Department of Radiation Oncology (Maastro), GROW Research Institute for Oncology and Reproduction, Maastricht University Medical Center+, Maastricht, the Netherlands

E-mail: o.glazunova@student.maastrichtuniversity.nl

**Abstract**

Objective: Radiotherapy uses precise doses of radiation to treat cancer, requiring accurate verification, e.g. using the Electronic Portal Imaging Device (EPID), to guide treatment. To develop an effective artificial intelligence (AI) model for error detection and treatment verification, a large and well-annotated dataset of EPID images is needed, however, acquiring such high quality real data is difficult. While synthetic EPID data could be a viable alternative, it is critical to ensure that this data is as realistic as possible to effectively train an accurate and reliable AI model. The measurement uncertainty that is not modeled in EPID predictions but is present on real measured EPID images can hinder downstream tasks such as error detection and classification. Our research aims to improve synthetic EPID data through image-to-image (I2I) translation based on deep generative modeling.
Approach: A dataset of 989 predicted EPID images and corresponding measured EPID images was used. We evaluate both paired and unpaired generative modeling approaches for this task. For the former, we introduce a novel modification of Variational Autoencoder (VAE) to I2I, a method that, to the best of our knowledge, has not been previously explored for this task. For the latter, we use UNsupervised Image-to-Image Translation Networks (UNIT).
Results: Our results show that both models achieved some degree of I2I translation, with our novel modification of the VAE model outperforming the UNIT model in improving key metrics (mean absolute error: 4.1 cGy vs 6.4 cGy; relative dose difference in-field: 2.5% vs 5.5%; absolute dose difference in-field: 5.3 cGy vs 10.8 cGy).
Significance: This enhanced synthetic data is expected to improve downstream tasks such as training neural networks for automated error detection and error classification in radiotherapy.

## 1. Introduction

Cancer poses a significant threat to global health, with a rising number of new cases and deaths reported each year. Approximately 18.1 million cancer cases were reported globally in 2020.[1] In the EU, there has been a 2.3% increase in new cancer cases in 2022 compared to 2020, with a total of 2.74 million cases this year, while cancer deaths rose by 2.4% compared to 2020.[2] Among cancer treatment options, radiation therapy (RT) stands out as the main approach for many cancer types, benefiting nearly half of all patients.[3] Its efficacy depends on the accuracy of the imaging and verification technologies that guide treatment and allow for accurate delivery of the radiation dose to the tumor while minimizing exposure to surrounding healthy tissues. Accurate imaging ensures that the radiation dose is delivered exactly as planned, maximizing treatment effectiveness and minimizing side effects.

One of the most important tools in RT is the electronic portal imaging device (EPID), which provides real-time dose images during treatment. These images are necessary to verify that the radiation doses are administered correctly. However, their evaluation causes a high workload and their interpretation is often difficult. In contrast, predicted EPID images generated by dedicated EPID software are easier to obtain, but often lack the realism of the real images as they fail to capture the uncertainties and noise of the actual dosimetry process in vivo, i.e., performed with a patient in the beam.[4–6] Lack of realism in simulated data (see Fig. 1) can hinder downstream tasks such as AI-based automated detection and classification of treatment errors.

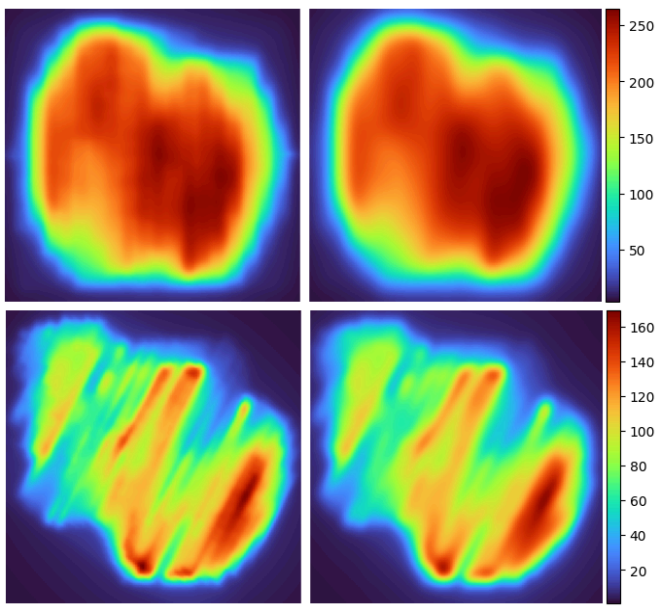

**Fig. 1**. EPID dose distribution image pairs: left - predicted (simulated), right - measured (real).

To address these issues, our research aims to improve the realism of simulated EPID data using deep generative models, capable of generating new images based on existing data. By capturing real-world artifacts and transferring them to simulated dose images with image-to-image (I2I) translation, we aim to transform predicted EPID images, making them closely resemble real measured dose images. This process involves using a dataset of paired predicted and measured open-field EPID images to train our models.

Ultimately, the enhanced synthetic data generated through our methods is expected to improve the performance of downstream tasks such as AI-based detection and classification of treatment errors.[6] This will help to ensure that RT is delivered more accurately and safely, ultimately improving patient outcomes.

**2. Methods**.

We implemented and evaluated two approaches to I2I translation: paired and unpaired. The paired approach was implemented via a Variational Autoencoder (VAE)[7–9] which learns to encode input images into latent space and decode them back to reconstruct target images. The unpaired approach was implemented via the UNsupervised Image-to-Image Translation Networks (UNIT)[10] framework, which uses both VAEs and Generative Adversarial Networks (GANs) to translate images from one domain to another without requiring paired training data. An innovative aspect of our study is the novel modification of VAE to enable I2I translation, a technique that, to the best of our knowledge, has not been previously explored with VAEs.

*2.1 Dataset*

The dataset used in this research consists of 1000 predicted-measured pairs of time-integrated EPID dose images in the form of matrices with original radiation dose values (in cGy) preserved. The predicted images are calculated based on treatment plans of 439 lung cancer patients; the measured images are obtained by delivering the treatment plan on the linear accelerator (linac). Both simulations and measurements were conducted in an open-field setting, i.e. without a patient or a phantom present in the radiation beam. The image (matrix) sizes vary depending on the image pair: some pairs are 384x512, while others are 512x512. Within each pair, the images have the same dimensions.

Several image pairs had to be excluded from the dataset due to various issues: value errors (exceptionally high dose values, e.g., 4 million cGy), shape or size mismatch (when predicted and measured radiation fields were drastically different based on visual inspection and mean absolute error calculation), and outliers (files of two patients who received a particular type of radiation therapy involving smaller fields and higher doses).

The resulting dataset used for training consisted of 989 pairs of single-channel images (435 unique patient numbers) with continuous (float) pixel values ranging from 0 to 688 cGy, representing predicted and measured radiation doses recorded by the EPID.

*2.1.1 Data Preprocessing.* Data preprocessing was performed on the fly and included cropping and size unification to prepare data for input into the models (Fig. 2). NaN values were replaced with 0.

Since the meaningful information about the radiation field occupies only a part of the image, pixels with values less than 5% of the maximum dose were removed. After that, size unification was performed to prepare images for input into the models: each cropped image was resized to 128x128 pixels.

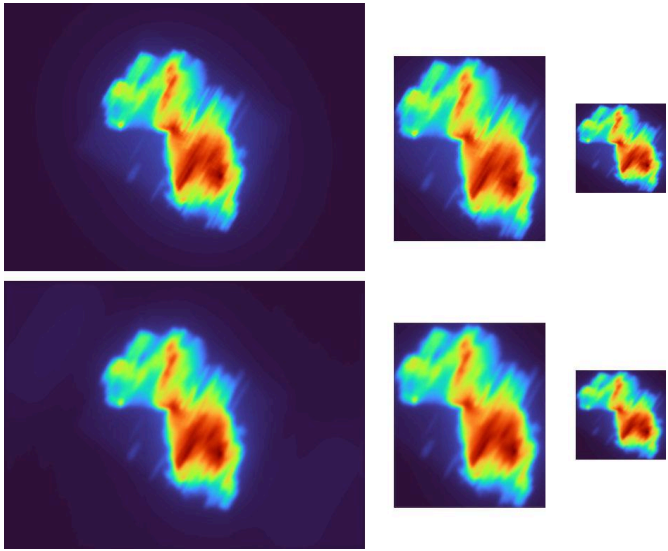

**Fig. 2**. Preprocessing step. From left to right: original image, crop, resize. First row: predicted image; second row: measured image.

To improve the generalization of the model and in an attempt to compensate for the small dataset size, data augmentation techniques were applied on the fly with specified probabilities to the input images: horizontal flip (probability 0.5), vertical flip (probability 0.5), and rotations (of 0, 90, 180, or 270 degrees; probability 1). The image values were normalized to the range [-1, 1] for training and testing, then restored to their original values for output visualization.

*2.2 Paired vs Unpaired I2I Translation*

Since the available dataset contains dose image pairs, both paired and unpaired I2I translations are possible.

In the paired approach, the model learns to match input images with corresponding target images, capturing direct relationships between each pair. This can assist the model in learning the nuances present in the pairs. The paired approach helps the model to learn meaningful differences between input and target more effectively. Research shows that unpaired generative models still benefit from using paired data as an auxiliary supervision to achieve better I2I translation results in medical imaging.[11] However, since our dataset size is quite small, the model can learn to remember specific mappings between input and target pairs, causing overfitting.

In the unpaired approach, the model learns to translate images from one domain to another by finding common features and patterns across the entire dataset. This method focuses on the general characteristics of images in each domain rather than on specific relationships in data pairs. Additionally, unpaired I2I translation can be considered a form of a regularization technique[12] that may help mitigate overfitting compared to paired I2I translation. Therefore, both paired and unpaired approaches were implemented and evaluated.

*2.3 Model Selection*

The goal of this research is to create synthetic data that is very similar to real measurement data. Deep generative modeling[13] was chosen for this purpose because it learns the underlying distribution of data and generates new samples based on that distribution. Deep generative models can produce realistic and diverse samples, making them ideal for tasks such as data synthesis,[14] which is important when dealing with limited real-world data.

Among various generative models, deep latent variable models (DLVMs) were selected due to their ability to efficiently capture complex data distributions and generate high-quality synthetic data. DLVMs, such as VAEs and GANs, extend traditional latent variable models by incorporating deep learning techniques that allow them to model complex patterns and dependencies in data. DLVMs can compactly represent data in a lower-dimensional latent space, generate new data samples from the latent space, and model complex, multimodal data distributions.

Among DLVMs, VAEs were chosen for several reasons:
- Being likelihood-based, VAEs allow the use of likelihood for uncertainty estimation; they can explicitly model the likelihood of the generated samples: the model assigns probability to its generated data matching the distribution of the training data, indicating how plausible the generated samples are.[9]
- VAEs have a relatively good explainability thanks to their interpretable latent space: by interpolating dimensions in the latent space, it is possible to understand which factors of variation have been learned, as well as their disentanglement level.[15]
- VAEs provide more stable training than other DLVMs, e.g., GANs, and possess more efficient sampling mechanisms.[16]
- VAEs are effective with limited training data.[8]

*2.4 Paired I2I Translation Implementation via Variational Autoencoder*

VAEs learn to encode data into a latent space and then decode it back to reconstruct the original data. The architecture consists of an encoder and a decoder network (Fig. 3). The encoder compresses the input data into a lower-dimensional latent space, and the decoder reconstructs this data by sampling from this latent space representation.

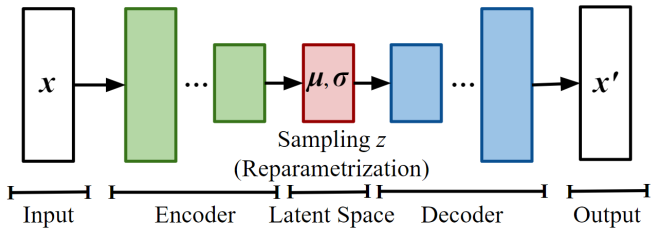

**Fig. 3**. VAE Architecture. $x$ - input, $x'$ - output, $z$ - latent variables, $\mu$ - mean, $\sigma$ - variance.

Unlike producing a single point in the latent space, VAEs generate a distribution, usually Gaussian, characterized by mean and variance. It is based on optimizing the likelihood of data that is marginalized over the auxiliary latent random variable. The likelihood is bound from below by evidence lower bound (ELBO) which allows us to utilize it in the optimization objective:

$$\text{ELBO} = \mathbb{E}_{q(z|x)}[\log p(x|z)] - D_{\text{KL}}[q(z|x)\|p(z)]$$

Here, $x$ refers to the input data, $z$ refers to the latent variables, which are a lower-dimensional representation of the input data $x$ in the latent space. The term $\mathbb{E}_{q(z|x)}[\log p(x|z)]$ represents the expected log-likelihood of the data given the latent variables, where the expectation is taken over the approximate posterior distribution $q(z|x)$. The second term, $D_{\text{KL}}[q(z|x)\|p(z)]$, is the Kullback-Leibler (KL) divergence between the approximate posterior $q(z|x)$ and the prior distribution $p(z)$. The ELBO aims to balance the reconstruction accuracy of the data with the regularization of the approximate posterior that should be close to the prior.[8]

The latent space represents an information bottleneck that compresses the flow of information and extracts the meaningful factors of variations.[8] The posterior distribution of the latent variables is approximated by optimizing a lower bound on the marginal likelihood using variational inference.

*2.4.1 Reparametrization Trick.* To enable backpropagation through the stochastic sampling process, VAEs use the reparameterization trick.[8,9] This involves sampling from a normal distribution using the mean ($\mu$) and square root of variance $\sigma^2$ obtained from the encoder:

$$z = \mu + \sigma \cdot \epsilon$$

where noise $\epsilon$ is sampled from the standard normal distribution $\mathcal{N}(0, I)$.

*2.4.2 Loss Function.* For the VAE optimization procedure, the negated estimate of ELBO is employed as a loss function. It consists of two components: the reconstruction loss and the KL divergence. The reconstruction loss measures how well the decoded output matches the input, while the KL divergence ensures that the learned posterior distribution is close to a standard normal.

$$\mathcal{L} = \mathcal{L}_{\text{rec}} + \mathcal{L}_{\text{KL}}$$

In the case of Gaussian log-likelihood, the reconstruction loss can be represented as the mean squared error (MSE) between the input $x$ and the reconstructed output $x'$:

$$\mathcal{L}_{\text{rec}} = \|x - x'\|^2$$

The KL divergence term ensures that the latent space distribution approximates a standard normal distribution. It is given by:

$$\mathcal{L}_{\text{KL}} = D_{\text{KL}}[q(z|x)\|p(z)]$$
$$= \frac{1}{2}\sum_{i=1}^{d}\left(1 + \log(\sigma_i^2) - \mu_i^2 - \sigma_i^2\right)$$

where $d$ is the dimensionality of the latent variable $z$, and $i$ is an index that runs over the dimensions of the latent variable $z$.

*2.4.3 Novel Approach: Substituting the VAE Reconstruction Term.* To the best of our knowledge, VAEs have not been previously used for I2I translation. We chose VAE for the paired approach due to its advantages in stable training, efficient sampling, and effective performance with limited training data. Additionally, we assumed that VAE's known tendency to produce blurry images could be beneficial in our task of transferring a degree of blurriness and uncertainty to the predicted EPID images.

In our new approach, we modified the standard VAE architecture to suit the aim of the research. Given that the predicted (input) and measured (target) images are very similar, we replaced the VAE reconstruction term with the target image $x_2$ instead of the input image $x_1$:

$$\mathcal{L}_{\text{rec}} = \|x_2 - x'\|^2$$

Our model learns features specific to the input image, compresses them into the latent space, and then reconstructs the image by comparing it with the target image (Fig. 3). This modification enables VAE to perform I2I translation instead of image reconstruction.

*2.4.4 Implementation.* The VAE model was implemented using PyTorch.[17] Since our VAE was constructed from scratch, i.e., no ready-made solution was adapted, a series of experiments were conducted to determine the optimal configuration.

For the research objective, the VAE's encoder and decoder were implemented as convolutional neural networks (CNNs) to capture the spatial structure of the input images. The encoder consists of five convolutional layers and two fully connected layers; the decoder consists of one fully connected layer and five transposed convolutional layers (a detailed description of the convolutional layers is provided in Appendix A).

The loss function used in our VAE implementation is based on the standard one for most VAEs, but includes an additional weighting factor $\beta$, making it a β-VAE.[15] A β-VAE is a variation of the standard VAE where the KL divergence term is weighted by a factor $\beta$ to control the balance between the reconstruction loss and the KL divergence:

$$\mathcal{L} = \mathcal{L}_{\text{rec}} + \beta \mathcal{L}_{\text{KL}}$$

In this implementation, if $\beta = 1$, it is a standard VAE; if $\beta \neq 1$, it is a β-VAE. The $\beta$ setting allows for a trade-off between latent representation capacity and an emphasis on the prior distribution, which can be beneficial in various applications.

*2.4.5 Training and Testing.* The VAE's training process optimizes the model using stochastic gradient descent. The model parameters are updated to minimize the total loss.

The training setup consisted of setting the model to the training mode and specifying the hyperparameters, including learning rate, batch size, number of epochs, and balancing weight $\beta$. Dimensions of the latent space were calculated dynamically based on input. For reproducibility, a random seed was introduced, to ensure that every experiment is conducted on the same set of samples and in identical conditions. After each epoch, the model's performance was evaluated on a validation set to check for overfitting.

The final settings of the VAE model were determined through experimentation and hyperparameter tuning using the Optuna framework: Input Dimension = 1 (single-channel images); Hidden Dimensions = [64, 128, 256, 512]; Batch Size = 64; Number of Epochs = 10000; Balancing weight ($\beta$) = 100; Adam optimizer with learning rate 1e-4; rectified linear unit (ReLU) activation function for intermediate layers, hyperbolic tangent (tanh) for the final layer of the decoder. After training, the model's performance was assessed via a set of metrics (explained in section 2.7) on a separate test set to evaluate how well the model generalizes to previously unseen data.

## 2.5 Unpaired I2I Translation Implementation via UNsupervised Image-to-Image Translation Networks

For the unpaired approach, a hybrid DLVM was employed: the UNIT (UNsupervised Image-to-Image Translation Networks) framework, developed by NVIDIA. It is specifically designed for unpaired I2I translation and incorporates both VAE and another type of DLVM - GAN.[18] Here, the learning problems of the VAE, GAN, and cycle-consistency constraints are jointly solved to perform I2I translation.

GANs take advantage of game theory methods: they consist of a generator and a discriminator engaged in adversarial training to refine image generation.[18] The generator, the neural network responsible for synthetic data creation, initially produces random output that may not resemble the target data. At the same time, another neural network, the discriminator, evaluates the generated and real data in an attempt to distinguish them by assigning high probabilities to the real data and low probabilities to the generated data. This adversarial learning process continues iteratively: the generator tries to fool the discriminator by maximizing the probability of misclassification, and the discriminator improves its ability to distinguish between real and generated data, eventually converging based on minimizing the loss function.

GANs are known for producing high-quality and realistic outputs, making them especially suitable for tasks like image synthesis and I2I translation. However, GANs require a high volume of training data and benefit greatly from data variability; their training speed is relatively slow[19]. The explainability is difficult because of the adversarial training process. The learning of GANs is based on bi-level optimization which is notoriously difficult to train: GANs are known for their delicate balance during training. If one network (generator or discriminator) becomes too dominant, it can disrupt the equilibrium needed for effective adversarial training. GANs are also sensitive to hyperparameters such as learning rates, architecture choices, and initialization.

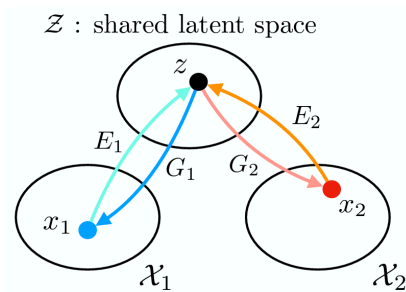

**Fig. 4**. The shared latent space assumption that UNIT is based on. $\mathcal{Z}$ - shared-latent space; $z$ - latent code; $X_1$, $X_2$ - different domains; $x_1$, $x_2$ - pair of corresponding images from different domains; $E_1$, $E_2$ - encoders; $G_1$, $G_2$ - generators. Reproduced from Liu et al.[10]

By combining both VAE and GAN, UNIT attempts to mitigate their drawbacks while combining their strengths: probabilistic approach and stable training of VAE and high-quality output images of GAN. It is based on the shared latent space assumption (Fig. 4), that a pair of corresponding images ($x_1$, $x_2$) in two different domains $X_1$ and $X_2$ can be mapped to the same latent code $z$ in a shared-latent space $Z$.

In terms of performance, UNIT's results are on par with GANs.[10] Among others, they produce results comparable to CycleGAN when performing I2I translation for magnetic resonance (MR) images.[20] UNIT can produce high-quality output like GANs while benefiting from the stable training of VAE.

*2.5.1 UNIT Architecture.* UNIT uses two autoencoders with a shared latent space to perform I2I translation between two domains. The adversarial training objective interacts with a weight-sharing constraint, which enforces a shared latent space, to generate corresponding output images in two domains, while the VAE relates translated images with input images in the respective domains[10].

The VAE component of UNIT aims to minimize a variational upper bound, ensuring that the latent code distribution closely matches the prior distribution. The VAE objective function includes the reconstruction loss and the KL divergence:

$$\mathcal{L}_{\text{VAE}} = \mathcal{L}_{\text{rec}} + \lambda_1 D_{\text{KL}}[q(z|x)\|p(z)]$$

where $\mathcal{L}_{\text{rec}}$ is the reconstruction loss, and $\lambda_1$ is a hyperparameter controlling the weight of the KL divergence term. The prior distribution is modeled as a zero-mean Gaussian:

$$p(z) = \mathcal{N}(z|0, I)$$

where $z$ represents the latent variable, and $I$ is the identity matrix.

The GAN component of UNIT ensures that the translated images resemble the target domain images. The objective functions for the GANs are conditional GAN objectives:

$$\mathcal{L}_{\text{GAN}} = \mathbb{E}_{x \sim p_{\text{data}}}[\log D(x)] + \mathbb{E}_{z \sim p(z)}[\log(1 - D(G(z)))]$$

where $D$ is the discriminator, and $G$ is the generator. The hyperparameter $\lambda_0$ controls the impact of the GAN objective functions.

The cycle-consistency constraint ensures that translating an image to the target domain and back to the original domain results in an image similar to the original. The VAE-like objective function models this constraint:

$$\mathcal{L}_{\text{cycle}} = \mathcal{L}_{\text{rec}} + \lambda_3 D_{\text{KL}}[q(z|x)\|p(z)] + \lambda_4 D_{\text{KL}}[q(z|G(z))\|p(z)]$$

where the negative log-likelihood objective term ensures that a twice-translated image resembles the input, and the KL divergence terms penalize deviations from the prior distribution. The hyperparameters $\lambda_3$ and $\lambda_4$ control the weights of the different objective terms.

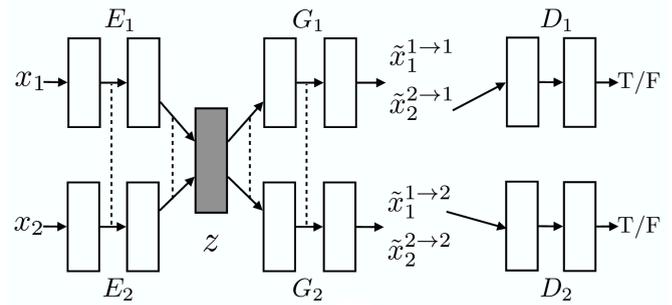

**Fig. 5**. The UNIT architecture. $z$ - latent space; $x_1$, $x_2$ - pair of corresponding images from different domains; $\tilde{x}_1^{1\to1}$, $\tilde{x}_2^{2\to2}$ - self-reconstructed images; $\tilde{x}_1^{1\to2}$, $\tilde{x}_2^{2\to1}$ - domain-translated images; $E_1$, $E_2$ - encoders; $G_1$, $G_2$ - generators; $D_1$, $D_2$ - discriminators. Reproduced from Liu et al.[10]

The UNIT training is based on solving a min-max problem where the optimization aims to find a saddle point. This can be seen as a two-player zero-sum game where the first player consists of the encoders and generators, and the second player consists of the adversarial discriminators. The optimization alternates between updating the discriminator and the encoder-generator networks (Fig. 5). During the gradient ascent phase, the discriminators $D_1, D_2$ are updated while keeping the encoders $E_1$, $E_2$ and generators $G_1$, $G_2$ fixed. Conversely, in the gradient descent phase, the encoders $E_1$, $E_2$ and generators $G_1$, $G_2$ are updated while keeping the discriminators $D_1$, $D_2$ fixed. This alternating gradient update ensures that the encoders and generators learn to produce realistic translations that fool the discriminators while maintaining cycle consistency and adhering to the latent space regularization.

*2.5.2 Impelementation.* Since UNIT is a ready-made framework designed specifically for unpaired I2I translation, minimal adjustments were required. We modified the original convolutional layers, implemented with PyTorch, to accept single-channel input instead of three-channel (RGB) images. The data loader was substituted by the one from our VAE implementation, also developed in PyTorch. Both VAE and UNIT use identical image transforms and share the same code for evaluation metrics and plotting functions, to ensure the correctness of the evaluation.

### 2.6 Experiments

The experiments were conducted on a system with CUDA support for GPU acceleration (NVIDIA TITAN V with 12 Gb memory). The models were trained for a specified number of epochs (VAE) or iterations (UNIT). After training, the models were evaluated on the same test set to generate samples and calculate the performance metrics. Training time was recorded as well (see Table 1).

For the paired approach, the dataset was divided into training, validation, and testing sets with a split ratio of 80%-10%-10%, respectively. As a result, 793 training samples, 96 validation, and 100 test samples were obtained. Patient numbers were ensured to be unique across all sets to ensure reliable model evaluation.

To ensure equal conditions of the experiments for both approaches, the same split with the same data was maintained. Since the UNIT framework is designed for unsupervised (unpaired) I2I translation, it does not include traditional validation steps during training, as opposed to supervised learning tasks where validation accuracy or loss is monitored. Therefore, the validation set was not used for the UNIT framework.

*2.7 Performance Evaluation*

Experiments were conducted to evaluate the performance of the chosen models, as well as compare paired and unpaired I2I translation approaches. Metrics for image similarity and dose accuracy were employed, as well as latent space evaluation using the K-means clustering algorithm.

*2.7.1 Image Similarity Metric.* Image similarity metrics are methods that produce a quantitative evaluation of the similarity between two or more images. The chosen metrics were selected due to their diverse characteristics and comprehensive coverage of various aspects of image quality and similarity.

**Mean Absolute Error (MAE)**: Measures the average absolute difference between predicted and measured values:
$$\mathrm{MAE} = \frac{1}{N} \sum_{i=1}^{N} |y_i - \hat{y}_i|$$
Here, $N$ is the number of pixels, $y_i$ is the ground truth value, and $\hat{y}_i$ is the predicted value.[21] There are no universal thresholds for MAE, but lower MAE indicates less deviation from true values, hence better image quality.

**Structural Similarity Index Measure (SSIM)**: Assesses the structural similarity between two images by considering luminance (intensity of the recorded object, i.e., the image's pixel values), contrast (difference/variation in luminance), and structure (correlation of the luminance of two images). A mathematical description of the SSIM is provided by Peng et al.[22]:
$$\mathrm{SSIM}(x,y) = \frac{(2\mu_x\mu_y + C_1)(2\sigma_{xy} + C_2)}{(\mu_x^2 + \mu_y^2 + C_1)(\sigma_x^2 + \sigma_y^2 + C_2)}$$
Here, $\mu_x$ and $\mu_y$ are the average pixel values, $\sigma_x^2$ and $\sigma_y^2$ are the variances, $\sigma_{xy}$ is the covariance, and $C_1$ and $C_2$ are constants to stabilize the division.[23] SSIM values range from -1 to 1, with values closer to 1 indicating higher similarity.

**Peak Signal-to-Noise Ratio (PSNR)**: Measures the ratio between the maximum possible power of a signal and the power of corrupting noise, expressed in decibels (dB):
$$\mathrm{PSNR} = 20 \log_{10}\left(\frac{\mathrm{MAX}_I}{\sqrt{\mathrm{MSE}}}\right)$$
Here, $MAX_I$ is the maximum possible pixel value of the image, and $MSE$ is the mean squared error between the predicted and ground truth images. Higher PSNR values indicate better image quality.

**Fréchet Inception Distance (FID)**: Estimates the similarity between generated and real data, typically in the context of GANs or other generative models. It measures the similarity between the distribution of generated images and the distribution of real images on a feature level, by measuring the Fréchet distance between their feature representations extracted from a pre-trained Inception-v3 model[24]. FID is calculated as follows:
$$\mathrm{FID} = \|\mu_r - \mu_g\|^2 + \mathrm{Tr}(\Sigma_r + \Sigma_g - 2(\Sigma_r\Sigma_g)^{1/2})$$
where $\mu_r$ and $\mu_g$ are the mean feature vectors over all real and generated images, respectively; $\Sigma_r$ and $\Sigma_g$ are the covariance matrices of the real and generated images, respectively; $\|\mu_r - \mu_g\|^2$ is the squared difference between the means; $\mathrm{Tr}(\Sigma_r + \Sigma_g - 2(\Sigma_r\Sigma_g)^{1/2})$ is the trace of the covariance matrices, adjusted for their similarity. Lower FID means that the generated images are more similar to the real images: a FID score close to 0 suggests that the generated images are nearly indistinguishable from the real images in terms of the statistical properties of their features.

*2.7.2 Dosimetric Accuracy Metrics.* These metrics evaluate the precision and reliability of dose distributions in radiotherapy and ensure that the predicted dose matches the measured dose. The dose difference analysis was conducted to evaluate the dose distribution accuracy in different regions of interest (ROIs). Based on the dose values, the regions were defined as follows:
- In-field: pixels with dose values ≥50% of the maximum dose of the target image.
- Out-of-field: pixels with dose values <50% of the maximum dose of the target image.

The absolute dose difference (ADD) is a direct measure of the absolute dose difference between the predicted and reference dose images. ADD per pixel was calculated as follows:
$$\mathrm{ADD} = |D_c - D_{\mathrm{ref}}|$$
where $D_c$ is the compared dose (in our case, input or output, depending on which pairs we compare), and $D_{\mathrm{ref}}$ is the reference dose (in our case, target or input). ADD gives the actual difference in dose units (in our case, cGy), which is useful for assessing the magnitude of the difference.

The relative dose difference (RDD) is a normalized measure that evaluates the dose difference relative to the maximum reference dose. RDD per pixel was calculated as follows:

$$\mathrm{RDD} = \frac{|D_c - D_{\mathrm{ref}}|}{D_{\mathrm{ref\_max}}} \times 100\%$$

where $D_c$ is the compared dose, $D_{\mathrm{ref}}$ is the reference dose, and $D_{\mathrm{ref\_max}}$ is the maximum value in the reference dose.[25] RDD provides a percentage difference, making it easier to compare across different dose levels and scales. For the evaluation, the mean ADD and the mean RDD were calculated for each ROI.

*2.7.3 Latent Space Evaluation.* The latent space is a lower-dimensional representation that can be used for clustering. Its visualization can show whether the model sees patterns in data: clustering indicates that the VAE has learned to group similar patterns, which is a sign of good representation learning.

First, we pass the dataset through the model's encoder to obtain the latent representations. These latent vectors serve as the features for clustering. Then we apply the K-means clustering algorithm to the latent vectors. K-means clustering aims to split $n$ observations into $K$ clusters, where each observation belongs to the cluster with the nearest mean. The objective function for K-means is:

$$\min_C \sum_{i=1}^{K} \sum_{z \in C_i} \|z - \mu_i\|^2$$

where $C = \{C_1, C_2, \ldots, C_K\}$ are the clusters, $z$ is a data point within the cluster $C_i$, and $\mu_i$ is the centroid of the cluster $C_i$, i.e., the mean of all data points $z$ in the cluster $C_i$.

Since we do not have prior knowledge about what clusters exist in our data, the number of clusters $K$ is chosen based on the silhouette score:

$$s(i) = \frac{b(i) - a(i)}{\max(a(i), b(i))}$$

where $a(i)$ is the mean distance between $i$ and all other points in the same cluster; $b(i)$ is the mean distance between $i$ and all points in the nearest cluster to which $i$ does not belong. The silhouette score measures how similar an object is to its cluster compared to other clusters. It ranges from -1 to +1, where a higher score (closer to +1) indicates better-defined clusters. The optimal $K$ maximizes the average silhouette score.

To visualize the clustered latent space, we employ the dimensionality reduction technique t-SNE.[26] The reduced 2D representations of the latent vectors can be plotted with different colors for each cluster.

The visualization of the latent space with K-means clustering helps in understanding how well the model has learned to disentangle the features of the input data. For example, clear separation between clusters indicates that the model has learned distinct features that differentiate between different classes or modes of the data, while overlapping clusters might suggest that the model has difficulty distinguishing between certain features. Compact clusters suggest that similar data points are mapped close to each other in the latent space. Spread-out clusters may indicate that the latent space representations for some classes are not well defined. Additionally, if labeled data is available (which is not the case in this study), coloring the clusters based on true labels can provide insights into the alignment between the latent space and the actual data distribution.

Table 1. Training time.

| Model | Training time |
|---|---|
| VAE | 22 hours 5 minutes 34 seconds |
| UNIT | 39 hours 38 minutes 50 seconds |

## 3. Results

*3.1 Metrics Comparison*

All metrics values are presented in the form of a barplot visualization provided in Fig. 6. Additionally, Appendix B contains a table with all metrics values, and Appendix C contains scatter plot comparisons for all metrics per test sample.

The "Inputs vs Target" metrics are identical for VAE and UNIT. These metrics were calculated separately for both models to ensure a fair comparison and to verify the accuracy of the calculations of the metrics. They represent the baseline quality that our models must surpass, helping us assess whether the models improve upon the baseline quality of existing predicted images by comparing the input (predicted images) to the target (measured images). The overall scores of the "Inputs vs Target" metrics presented in Fig. 6 and in Appendix B indicate a high level of similarity between predicted and measured images, with a low MAE, high PSNR, and high SSIM values, demonstrating that our input is already quite close to the target.

The "Output vs Target" metrics show how close the images generated by our models are to the real measured images. The goal of this study is to achieve better performance for this comparison than for "Input vs. Target". The "Output vs. Input" metrics show how much the generated images by the AI models differ from the original predicted images. On the one hand, these values help assess whether the models were able to effectively grasp and learn the essential features of the input images. On the other hand, by comparing "Output vs Input" with "Output vs Target", we can determine whether the changes made by the models result in a closer match to the target images, thus validating the effectiveness of our I2I translation approach.

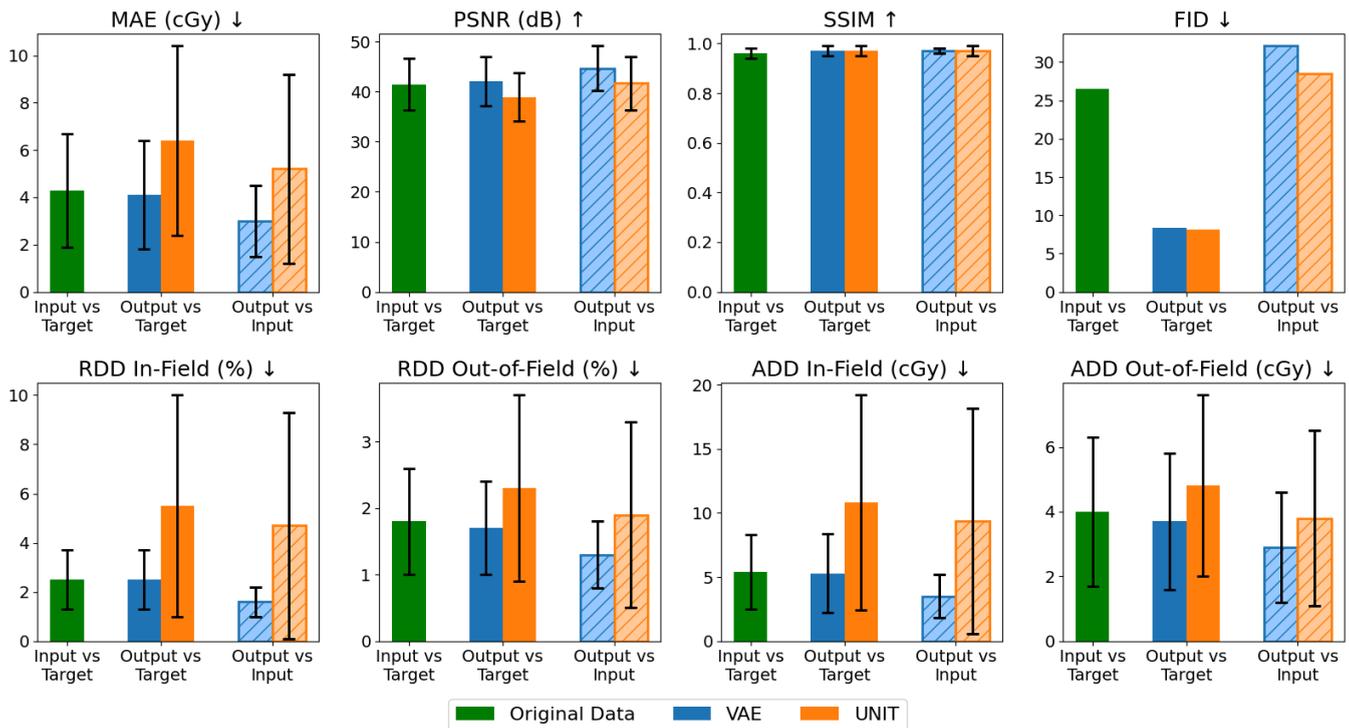

**Fig. 6.**. Image similarity and dose accuracy metrics for VAE and UNIT.

All metrics, except for FID, were calculated as means of values received for individual image pairs in the test set. FID is the only metric that we employed which can be calculated for the dataset as a whole. In the case of VAE, the scores we obtain, while being consistent in different runs for Input vs Target, may slightly fluctuate for comparisons with the output (within one-tenth of a decimal place), despite our measures taken to ensure the identical experimental conditions, such as maintaining the same set of test samples and introducing a fixed random seed. These fluctuations are also present in the case of UNIT, albeit less frequently and on a smaller scale (within one-hundredth of a decimal place).

*3.1.1 VAE Metrics Analysis.* In the case of VAE, the MAE for Output vs Target (4.1 ± 2.3 cGy) is slightly lower than Input vs Target (4.3 ± 2.4 cGy), indicating that the accuracy was preserved, rather than improved. The low MAE score for Output vs Input (3.0 ± 1.5 cGy) shows the model's ability to produce consistent outputs relative to the inputs, meaning that the transformations applied by the model preserve the essential features of the input images but do not make the output closer to the target than to the input.

The PSNR for Output vs Target (42.14 ± 4.91 dB) is slightly higher than Input vs Target (41.49 ± 5.17 dB), indicating that image quality didn't decrease post-translation. The PSNR for Output vs Input (44.71 ± 4.45 dB) is even higher, reflecting the consistency of the output images compared to the inputs.

The SSIM for Output vs Target (0.97 ± 0.02) is slightly higher than Input vs Target (0.96 ± 0.02), indicating a slight improvement in structural similarity after translation. An equally high SSIM score for Output vs Input (0.97 ± 0.01) shows that the output images are structurally similar to the input images.

The FID score for Output vs Target is 8.40. The FID score for Output vs Input is considerably higher at 31.66, indicating that the outputs are quite different from the inputs, which reflects substantial changes introduced by I2I translation and also signifies that the output is closer to the target than to the input. The FID score for Input vs Target is 26.53, which is higher than Output vs Target, suggesting that I2I translation has made the outputs more similar to the targets compared to the inputs.

Both RDD and ADD values for Output vs Target are slightly lower than for Input vs Target. This indicates that the dosimetric accuracy has been preserved in all regions.

*3.1.2 UNIT Metrics Analysis.* The MAE for Output vs Target (6.4 ± 4.0 cGy) is higher than Input vs Target (4.3 ± 2.4 cGy), indicating that I2I translation not only did not reduce the uncertainties but also introduced additional discrepancies. The PSNR of Output vs Target (38.89 ± 4.82 dB) is lower than the Input vs Target image (41.49 ± 5.17 dB), indicating that the quality of the output images has decreased. This suggests that I2I translation did not improve the image quality when compared with the target image. The higher PSNR for Output vs Input (41.67 ± 5.32 dB) means that the output data is closer to the input data, maintaining consistency but not necessarily improving quality compared to the target.

The SSIM for Output vs Target (0.97 ± 0.02) is slightly higher than Input vs Target (0.96 ± 0.02), indicating a small improvement in the already very high structural similarity. SSIM for Output vs Input (0.97 ± 0.02) shows that the output images have the same structural similarity as both the input and the target images.

The FID score of Output vs Target is 8.16, indicating that the output images are very similar to the target images, when compared to Input vs Target and Output vs Input. It is much lower than the FID score of Output vs Input (28.50), indicating significant changes introduced by I2I translation. The FID score for Input vs Target is 26.53, which is higher than that for Output vs Target, suggesting that the I2I translation process made the output more similar to the target compared to the input, and therefore, yielded the desired result.

All RDD and ADD scores are higher for Output vs Target than for Input vs Target. It indicates that I2I translation has increased discrepancies in the dose distributions.

## 3.2 Image Results

*3.2.1 Visualization Methodology.* To visualize generated samples, two methods were used: concatenated comparison plots and dose profiles. The concatenated comparison plots show the input, output, and target images with the "turbo" colormap applied, together with a colorbar common for all images. The dose profiles were extracted from the row (cross-line) and column (in-line) of images containing the maximum dose, with an inset image of the output providing a visual reference of the dose distribution.

*3.2.2 Examples Selection.* The best and the worst cases for both models are presented. To determine the best and the worst samples, we normalized all metric values to range [0, 1]. For metrics where lower values indicate better performance (MAE, RDD, ADD), the normalized values were inverted by subtracting them from 1. A composite score for each output image was calculated by averaging the adjusted normalized scores. The image with the highest cumulative score was selected as the best case; the image with the lowest cumulative score was selected as the worst case. The selection was then verified by visual assessment of the samples and by manual metrics comparison. We also checked for outliers - images that scored exceptionally good or bad on some metric. However, all metrics showed consistent performance and did not contradict each other.

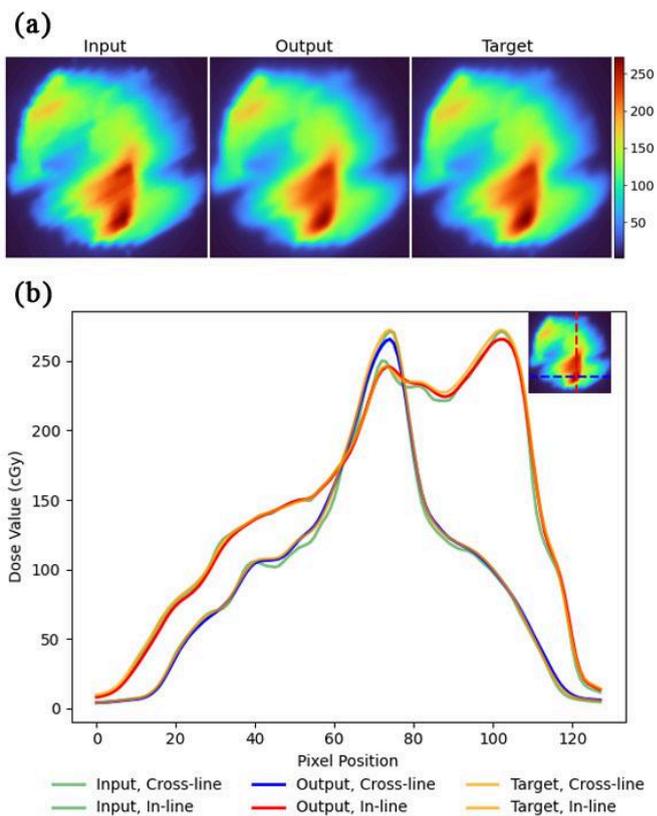

**Fig. 7**. The best sample of VAE. (a) Input, Output and Target. (b) Dose profiles.

**Table 2**. VAE best sample metrics.

| Metric | Input vs Target | Output vs Target | Output vs Input |
|---|---|---|---|
| MAE (cGy) | 2.0 | 1.7 | 2.1 |
| PSNR (dB) | 47.50 | 49.15 | 46.78 |
| SSIM | 0.99 | 0.99 | 0.98 |
| **RDD (%)**: | | | |
| In-field | 1.2 | 0.8 | 1.1 |
| Out-of-field | 0.6 | 0.6 | 0.7 |
| **ADD (cGy)**: | | | |
| In-field | 3.2 | 2.3 | 3.0 |
| Out-of-field | 1.7 | 1.6 | 2.0 |

*3.2.2 Best Cases.* The best test sample happened to be the same for VAE (Fig. 7, Table 2) and UNIT (Fig. 8, Table 3). When comparing the models' outputs, we observe that for the best samples, VAE and UNIT demonstrate similar performance, with VAE yielding slightly better results.

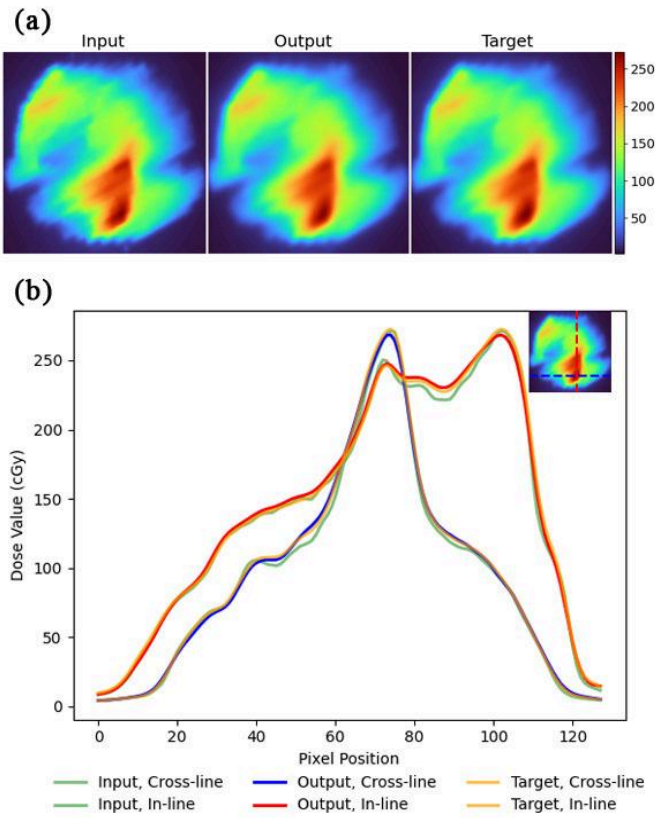

**Fig. 8**. The best samples of UNIT. (a) Input, Output and Target. (b) Dose profile.

**Table 3**. UNIT best sample metrics.

| Metric | Input vs Target | Output vs Target | Output vs Input |
|---|---|---|---|
| MAE (cGy) | 2.0 | 1.7 | 2.5 |
| PSNR (dB) | 47.50 | 49.23 | 45.89 |
| SSIM | 0.99 | 1.00 | 0.99 |
| **RDD (%)**: | | | |
| In-field | 1.2 | 1.0 | 1.5 |
| Out-of-field | 0.6 | 0.6 | 0.8 |
| **ADD (cGy)**: | | | |
| In-field | 3.2 | 2.6 | 4.0 |
| Out-of-field | 1.7 | 1.5 | 2.2 |

*3.1.3 Worst Cases.* The results show that VAE and UNIT struggle with different kinds of dose distribution images (Fig. 9, Table 4 for VAE; Fig. 10, Table 5 for UNIT). In general, on the bad samples, UNIT performs significantly worse than VAE.

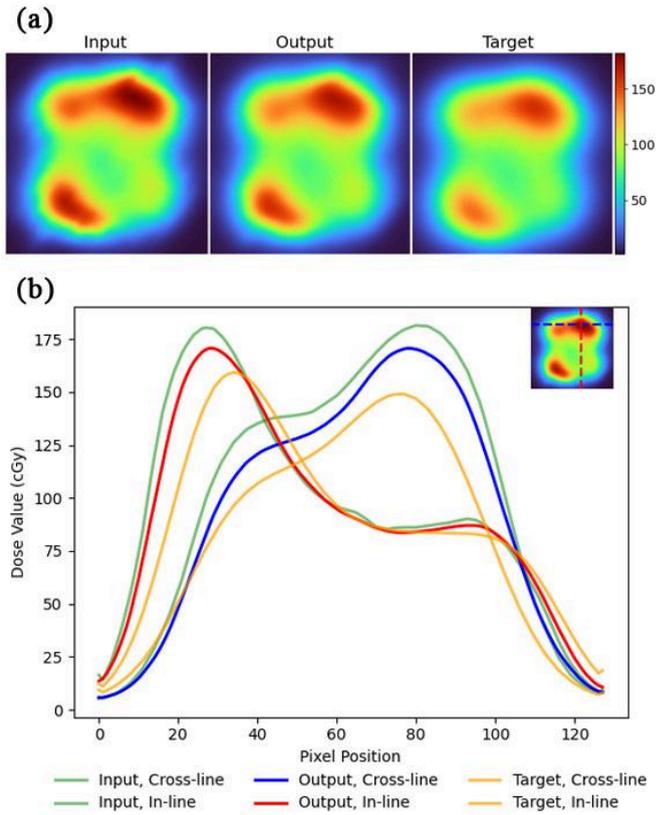

**Fig. 9**. One of the worst samples of VAE. (a) Input, Output and Target. (b) Dose profile.

**Table 4**. VAE worst sample metrics.

| Metric | Input vs Target | Output vs Target | Output vs Input |
|---|---|---|---|
| MAE (cGy) | 10.6 | 8.5 | 3.8 |
| PSNR (dB) | 33.13 | 35.53 | 42.27 |
| SSIM | 0.87 | 0.92 | 0.98 |
| **RDD (%)**: | | | |
| In-field | 9.2 | 6.5 | 3.4 |
| Out-of-field | 5.2 | 4.6 | 1.5 |
| **ADD (cGy)**: | | | |
| In-field | 14.7 | 10.4 | 6.1 |
| Out-of-field | 8.3 | 7.4 | 2.8 |

It is important to note that even though this is VAE's worst sample, all metric values are still improved when comparing Input vs Target with Output vs Target.

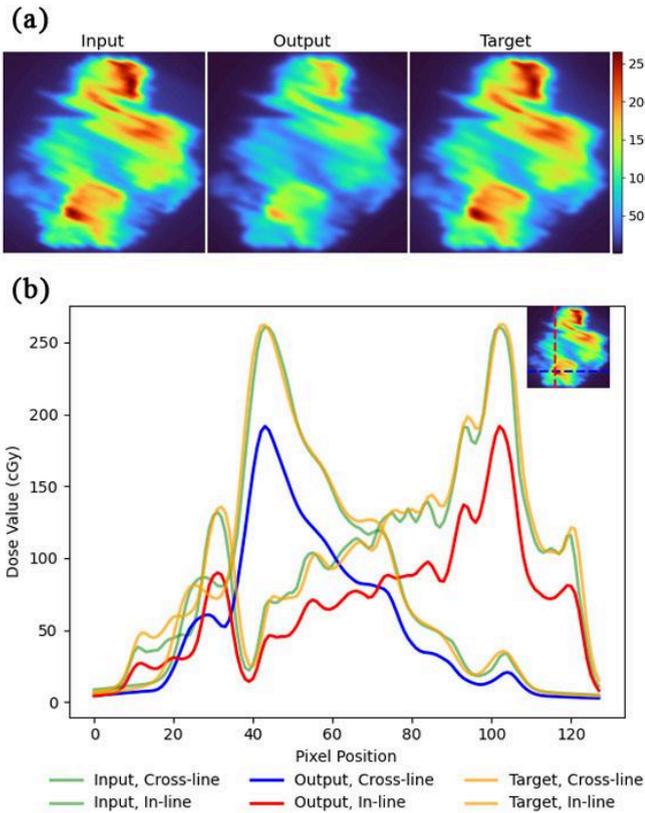

**Fig. 10**. The worst sample of UNIT. (a) Input, Output and Target. (b) Dose profile.

Table 5. UNIT worst sample.

| Metric | Input vs Target | Output vs Target | Output vs Input |
|---|---|---|---|
| MAE (cGy) | 4.9 | 22.6 | 21.6 |
| PSNR (dB) | 38.87 | 27.02 | 27.65 |
| SSIM | 0.94 | 0.87 | 0.88 |
| **RDD (%)**: | | | |
| In-field | 3.1 | 20.1 | 18.8 |
| Out-of-field | 1.5 | 5.3 | 5.3 |
| **ADD (cGy)**: | | | |
| In-field | 8.1 | 52.8 | 50.0 |
| Out-of-field | 4.1 | 14.0 | 14.2 |

We noticed UNIT's tendency to elevate dose values, especially in the case of uniformly shaped dose distributions with low values (Fig. 11). VAE performs better on such images, but also struggles with them: it tends to generate non-uniform patterns for large uniform low-dose areas.

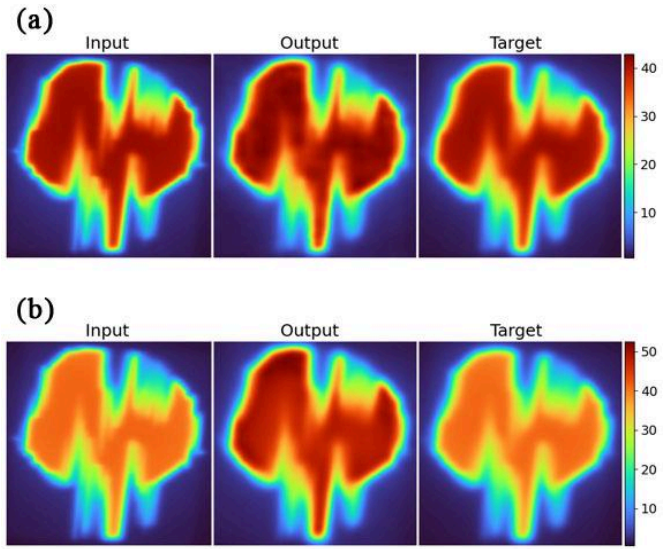

**Fig. 11**. Low-dose uniform sample. (a) VAE. (b) UNIT.

### 3.3 Latent Space Evaluation

When analyzing latent space cluster visualization, it is important to remember that the t-SNE dimensionality reduction algorithm was employed, so on these plots, there are many dimensions visualized as 2-dimensional space. t-SNE was applied to the latent variables sampled from the posterior distribution predicted by our VAE modification and by the VAE component of the UNIT framework. These latent variables capture the essential statistical properties of the input data. The X-axis presents the first dimension of the transformed latent space after dimensionality reduction using t-SNE; it aims to capture as much of the variance in the data as possible. The Y-axis is the second dimension after dimensionality reduction.

The number of clusters found by the silhouette score for the VAE and UNIT models is the same (three), suggesting a small number of categories in the data. In both plots, two clusters are relatively small and more compact, while one cluster is larger and more spread out. The similarity of the latent spaces constructed by two independent models may indicate that there are indeed three distinct categories of images in the dataset, one larger and more varied, and the other two smaller and possessing more uniform features.

In the case of VAE, clusters are more or less separated, with minor overlaps, suggesting that the model has learned to differentiate between categories (Fig. 12). The points within each cluster are relatively dense, indicating well-defined categories. The spread within the larger cluster suggests it might encompass more diverse features. In the case of UNIT, the latent space has more spread-out clusters with less distinct boundaries. (Fig. 13).

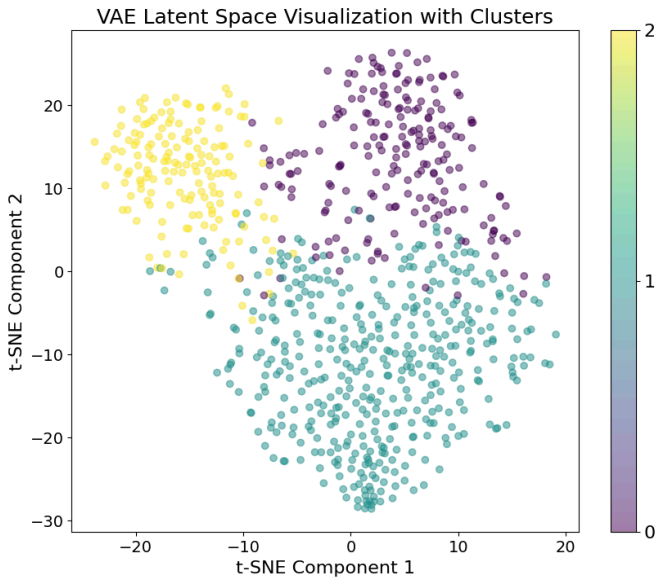

**Fig. 12**. The resulting latent space of VAE.

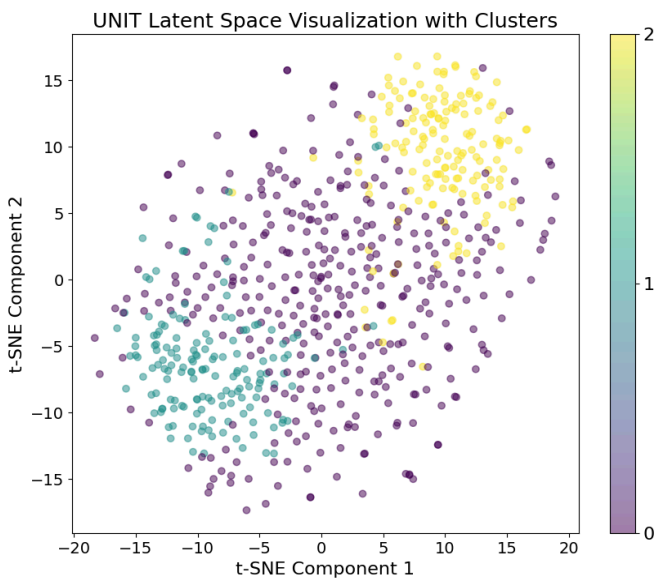

**Fig. 13**. The resulting latent space of UNIT.

## 4. Discussion

### 4.1 Results

The "Inputs vs Target" metrics presented in Fig. 6 and in Appendix B indicate that our input is very similar to the target already. This makes the task of I2I translation challenging for the models, as they must surpass the already high baseline metrics.

The VAE model shows slight improvements in MAE and PSNR, indicating a small reduction of uncertainties and better image quality. The slight improvement of the already very high SSIM means that structural integrity is maintained rather than considerably improved. The ADD and RDD suggest an improvement in the dosimetric accuracy. The FID score indicates effective I2I translation: the FID score for "Output vs Target" is much lower than for both "Input vs Target" and "Output vs Input." This confirms that, both for a human interpreter and for an AI, the output is perceptually closer to the target than to the input, signifying the successful completion of the I2I translation task

The UNIT model shows a slight improvement in SSIM and a considerable decrease in FID, suggesting better structural and perceptual similarity when compared to target images. Additionally, the FID scores indicate success in I2I translation, the same as in the case of VAE. However, the model output also displays increased MAE, decreased PSNR, and higher RDD and ADD values, indicating worsened overall image quality and dosimetric accuracy. The model's changes introduce more errors and inconsistencies, raising concerns about its effectiveness in practical applications.

Despite having a higher FID than UNIT, the VAE model performs better overall due to improved accuracy, consistency, and quality. The latent space analysis supports the metrics results, with VAE's superior performance being reflected in the more pronounced latent space structure. This can be attributed to VAE benefiting from the paired approach by having more information for training, while UNIT, despite being unpaired, also performed relatively well, considering the small dataset size. In summary, while both models achieve some degree of I2I translation, the VAE model shows more promise in practical application, making it a more robust choice for downstream applications.

Interestingly, the perceptual quality of the images improves for both VAE and UNIT, as evidenced by the low FID scores, whereas the traditional image quality metrics (MAE, SSIM, PSNR) and dosimetric accuracy metrics improve only for VAE. This suggests that the improvements seen from an AI-based perspective (FID) do not necessarily translate to better mathematical accuracy or physical accuracy in the context of EPID images, showing the gap between perceptual quality and practical accuracy.

Additionally, FID is the only metric that shows the best scores for "Output vs Target", not for "Output vs Input". This means that according to the FID metric, the translated images (outputs) are more similar to the target images than they are to the input images, demonstrating that from an AI-based perspective, a successful I2I translation has been achieved.

### 4.3 Comparison with Previous Work

Our study introduces a novel modification of VAE for I2I translation. Previous work in this area has mainly focused on GANs and CycleGANs for I2I translation tasks. For example, the work of Isola et al. demonstrated the effectiveness of GANs in paired I2I translation.[27] Additionally, Zhu et al. introduced CycleGAN, which allows for unpaired I2I translation by using cycle-consistency loss to enforce the similarity between input and output.[28] However, a disadvantage of GANs is that their training can be unstable[18].

In comparison, our VAE and UNIT approaches benefitted from stable training. In previous work, the UNIT framework, combining VAE and GAN, showed results comparable to other state-of-the-art methods but with added stability due to the VAE component.[20] For a similar purpose, Bao et al. proposed the use of Conditional Variational Autoencoder Generative Adversarial Network (CVAE-GAN) for multi-domain image generation, combining VAE and GAN to enhance stability and image quality.[29] Similarly, Huang et al. introduced Multimodal UNsupervised Image-to-image Translation (MUNIT), a framework that performs multi-modal I2I translation using a combination of VAE and GAN.[30] Additionally, recent work by Lee et al. with Diverse Image-to-Image Translation (DRIT), which also integrates disentangled representation learning into the I2I translation process, supports the notion that combining VAE with GAN components can yield stable and high-quality image translations.[31] Our results are consistent with these studies, demonstrating that VAE and GAN integration ensures stable training.

The superior performance of VAE in our research, both in terms of image quality and dosimetric accuracy, can perhaps be attributed to the paired approach being better suited for limited data, as well as VAE performing better in scenarios with limited training data.[8] We can conclude that our novel modification of VAE for I2I translation is a viable alternative to traditional GAN-based approaches.

*4.4 Metrics Discussion*

While the evaluation metrics employed are widely accepted and provide valuable information about model performance, they also have certain limitations and drawbacks. Therefore, a set of varied metrics had to be implemented to thoroughly evaluate the models' performance. This includes traditional image similarity metrics (MAE, SSIM, PSNR), dose accuracy metrics, and an AI-based metric that evaluates perceptual image quality (FID).

*4.4.1 Fréchet Inception Distance.* FID is widely used for evaluating the quality of generated images by comparing the distribution of generated images to real images. However, FID can be biased, leading to inaccurate evaluations under certain conditions, particularly when dealing with small datasets or assessing unpaired images.[32] Despite its shortcomings, FID remains one of the few metrics that evaluate the entire image distribution rather than individual image pairs. As an alternative to FID, future work could use Maximum Mean Discrepancy (MMD)[34] - a kernel-based statistical test used to measure the distance between two distributions, or Kernel Inception Distance (KID)[33], which also measures the similarity between the distributions of generated images and real images using features extracted from a pre-trained Inception model. KID was designed to be a more reliable alternative to FID with less susceptibility to bias, particularly in scenarios involving small sample sizes. Both MMD and KID evaluate perceptual image quality by comparing the feature distributions of real and generated images. Implementing them along with FID would help determine if FID is biased in our case.

*4.5 Future Work*

Future work should focus on training VAE and UNIT models on in vivo data, as open-field data may not reflect certain nuances present in real-world clinical settings. Investigating the application of both models to in vivo data is necessary to assess whether synthetic EPID data are sufficiently realistic. This transition may require additional model tuning due to increased data variability and visual artifacts. However, this increased variability can also provide our models with clearer information about the underlying distribution of uncertainties in measured EPID images, potentially improving the performance.

The next step involves applying the results of these models to downstream tasks such as AI-based error detection and classification. By comparing the performance of AI models trained on enhanced predicted EPID images with those trained on original predicted EPID images, we can evaluate if there is an improvement in accuracy and robustness when applied to real EPID measurements.

Additionally, given the promising results of the novel modification of VAE for I2I translation, further exploration of this method may be worthwhile. Since the latent space evaluation showed the presence of three clusters in the data, thereby performing an unsupervised classification task, it is possible to use this information to implement a Conditional Variational Autoencoder (CVAE).[35] CVAE is an extension of the standard VAE that includes additional information in the form of a condition to control the generation process. This additional information may be any form of auxiliary data such as class labels, attributes, or any relevant metadata. Introducing CVAE by conditioning on the latent space cluster structure can potentially improve the clustering and thus the overall performance of the model.

*4.6 Other Prospective Approaches*

If the results of this research prove to be insufficient to generate realistic images after applying them to in vivo data, it may be worth exploring other approaches to the task.

*4.6.1 De-Blur instead of Blur.* Currently, we transfer blur and uncertainties from the measured domain to the predicted domain. It would be interesting to evaluate the reverse process of de-blurring and cleaning the measured images.

Being an unpaired method, UNIT has been already trained to transfer styles from measured to predicted and vice versa. Training VAE to transfer style from the measured to the predicted domain and comparing the results with UNIT can provide valuable insights and help to evaluate this reversed approach. Moreover, the task of image deblurring is quite in demand, and many ready-made solutions exist, among them those that deal specifically with medical images, e.g., Denoising Convolutional Neural Network (DnCNN)[36] or DeeplabV3+.[37]

*4.6.2 Kernel Blur.* Since our primary goal is to introduce blur and uncertainties, traditional physics-based methods for dealing with blur may be of use. Among them, kernel blur is especially worth exploring. Kernel blur methods have been effectively used in astrophysics for many years to de-blur images of celestial objects.[38,39] Tran et al. adapted this approach to the field of AI, introducing blur operators from a dataset of sharp-blur image pairs into a blur kernel space.[40] The alternating optimization algorithm for blind image deblurring approximates an unseen blur operator by a kernel in the encoded space and searches for the corresponding sharp image. This method can be used both for image deblurring and blur synthesis by transferring existing blur operators from a given dataset into a new domain.

## 5. Conclusion

This study demonstrates that a novel VAE modification for I2I translation shows potential in improving the realism of synthetic EPID images. The VAE model maintained structural integrity and showed slight improvements in image quality and dosimetric accuracy compared to the UNIT model. These results indicate that the enhanced synthetic data generated by our modified VAE may be beneficial for downstream tasks such as AI-based error detection and classification, which will ultimately contribute to more accurate and reliable patient treatment.

## Acknowledgements

This study was partially funded by Varian, a Siemens Healthineers company (Project VeriDash).

## References

1. Ferlay J, Ervik M, Lam F, Colombet M, Mery L. International agency for research on cancer 2020. Glob Cancer Obs Cancer.

2. European Cancer Information System. Cancer Factsheets in EU-27 countries - 2022 [Internet]. 2023. Available from: https://ecis.jrc.ec.europa.eu/factsheets_2022.php

3. Otto K. Volumetric modulated arc therapy: IMRT in a single gantry arc. Med Phys. 2008 Jan;35(1):310–7.

4. Baeza JA, Wolfs CJA, Nijsten SMJJG, Verhaegen F. Validation and uncertainty analysis of a pre-treatment 2D dose prediction model. Phys Med Biol. 2018 Feb 2;63(3):035033.

5. Lin MH, Price RA Jr, Li J, Kang S, Li J, Ma CM. Investigation of pulsed IMRT and VMAT for re-irradiation treatments: dosimetric and delivery feasibilities. Phys Med Biol. 2013 Nov 21;58(22):8179–96.

6. Wolfs C. Quantitative methods for improved error detection in dose-guided radiotherapy [Internet]. University of Maastricht; 2020. Available from: https://cris.maastrichtuniversity.nl/files/71677582/c6780_embargo.pdf

7. Kingma DP, Rezende DJ, Mohamed S, Welling M. Semi-Supervised Learning with Deep Generative Models [Internet]. arXiv [cs.LG]. 2014. Available from: http://arxiv.org/abs/1406.5298

8. Kingma DP, Welling M. Auto-Encoding Variational Bayes [Internet]. arXiv [stat.ML]. 2013. Available from: http://arxiv.org/abs/1312.6114

9. Kingma DP, Welling M. An Introduction to Variational Autoencoders. Foundations and Trends® in Machine Learning. 2019;12(4):307–92.

10. Liu MY, Breuel T, Kautz J. Unsupervised Image-to-Image Translation Networks [Internet]. arXiv [cs.CV]. 2017. Available from: http://arxiv.org/abs/1703.00848

11. Paavilainen P, Akram SU, Kannala J. Bridging the Gap Between Paired and Unpaired Medical Image Translation. In: Deep Generative Models, and Data Augmentation, Labelling, and Imperfections. Springer International Publishing; 2021. p. 35–44.

12. Domingos P. A few useful things to know about machine learning. Commun ACM. 2012 Oct 1;55(10):78–87.

13. Tomczak JM. Deep Generative Modeling. Springer International Publishing; 18 p.

14. Bond-Taylor S, Leach A, Long Y, Willcocks CG. Deep Generative Modelling: A Comparative Review of VAEs, GANs, Normalizing Flows, Energy-Based and Autoregressive Models. IEEE Trans Pattern Anal Mach Intell. 2022 Nov;44(11):7327–47.

15. Higgins I, Matthey L, Pal A, Burgess C, Glorot X, Botvinick M, et al. beta-VAE: Learning Basic Visual Concepts with a Constrained Variational Framework. In: 5th International Conference on Learning Representations, ICLR 2017, Toulon, France, April 24-26, 2017, Conference Track Proceedings [Internet]. OpenReview.net; 2017. Available from: https://openreview.net/forum?id=Sy2fzU9gl

16. Ghosh P, Sajjadi MS, Vergari A, Black M, Scholkopf B. From variational to deterministic autoencoders. In: International Conference on Learning Representations. 2019.

17. Paszke A, Gross S, Massa F, Lerer A, Bradbury J, Chanan G, et al. PyTorch: An imperative style, high-performance deep learning library. Adv Neural Inf Process Syst [Internet]. 2019 Dec 3;abs/1912.01703. Available from: https://proceedings.neurips.cc/paper/2019/hash/bdbca288fee7f

92f2bfa9f7012727740-Abstract.html

18. Goodfellow IJ, Pouget-Abadie J, Mirza M, Xu B, Warde-Farley D, Ozair S, et al. Generative Adversarial Networks [Internet]. arXiv [stat.ML]. 2014. Available from: http://arxiv.org/abs/1406.2661

19. El-Kaddoury M, Mahmoudi A, Himmi MM. Deep Generative Models for Image Generation: A Practical Comparison Between Variational Autoencoders and Generative Adversarial Networks. In: Mobile, Secure, and Programmable Networking. Springer International Publishing; 2019. p. 1–8.

20. Welander P, Karlsson S, Eklund A. Generative Adversarial Networks for Image-to-Image Translation on Multi-Contrast MR Images - A Comparison of CycleGAN and UNIT [Internet]. arXiv [cs.CV]. 2018. Available from: http://arxiv.org/abs/1806.07777

21. Willmott C. On the validation of models. Phys Geogr. 1981 Jul 1;2:184–94.

22. Peng J, Shi C, Laugeman E, Hu W, Zhang Z, Mutic S, et al. Implementation of the structural SIMilarity (SSIM) index as a quantitative evaluation tool for dose distribution error detection. Med Phys. 2020 Apr;47(4):1907–19.

23. Wang Z, Bovik AC, Sheikh HR, Simoncelli EP. Image quality assessment: from error visibility to structural similarity. IEEE Trans Image Process. 2004 Apr;13(4):600–12.

24. Heusel M, Ramsauer H, Unterthiner T, Nessler B, Hochreiter S. GANs Trained by a Two Time-Scale Update Rule Converge to a Local Nash Equilibrium. In: Advances in Neural Information Processing Systems (NeurIPS). 2017.

25. Wolfs CJA, Verhaegen F. What is the optimal input information for deep learning-based pre-treatment error identification in radiotherapy? Phys Imaging Radiat Oncol. 2022 Oct;24:14–20.

26. Maaten L, Hinton GE. Visualizing Data using t-SNE. J Mach Learn Res. 2008;9:2579–605.

27. Isola P, Zhu JY, Zhou T, Efros AA. Image-to-image translation with conditional adversarial networks. Proc IEEE Comput Soc Conf Comput Vis Pattern Recognit. 2016 Nov 21;5967–76.

28. Zhu JY, Park T, Isola P, Efros AA. Unpaired image-to-image translation using cycle-consistent adversarial networks. In: Proceedings of the IEEE international conference on computer vision. 2017. p. 2223–32.

29. Bao J, Chen D, Wen F, Li H, Hua G. CVAE-GAN: Fine-grained image generation through asymmetric training. ICCV. 2017 Mar 29;2764–73.

30. Huang X, Liu MY, Belongie S, Kautz J. Multimodal Unsupervised Image-to-Image Translation [Internet]. arXiv [cs.CV]. 2018. Available from: http://arxiv.org/abs/1804.04732

31. Lee HY, Tseng HY, Mao Q, Huang JB, Lu YD, Singh M, et al. DRIT++: Diverse image-to-image translation via disentangled representations. Int J Comput Vis. 2020 Nov;128(10-11):2402–17.

32. Chong MJ, Forsyth D. Effectively Unbiased FID and Inception Score and where to find them [Internet]. arXiv [cs.CV]. 2019. Available from: http://arxiv.org/abs/1911.07023

33. Sutherland JD, Arbel M, Gretton A. Demystifying MMD GANs. In: International Conference for Learning Representations. 2018. p. 1–36.

34. Gretton A, Borgwardt KM, Rasch M, Scholkopf B, Smola A. A Kernel Two-Sample Test. J Mach Learn Res. 2012 Mar 1;13:723–73.

35. Sohn K, Lee H, Yan X. Learning structured output representation using deep conditional generative models. Adv Neural Inf Process Syst. 2015 Dec 7;3483–91.

36. Zhang K, Zuo W, Chen Y, Meng D, Zhang L. Beyond a Gaussian Denoiser: Residual Learning of Deep CNN for Image Denoising. IEEE Trans Image Process. 2017 Jul;26(7):3142–55.

37. Chen LC, Zhu Y, Papandreou G, Schroff F, Adam H. Encoder-decoder with atrous separable convolution for semantic image segmentation. ECCV. 2018 Feb 7;833–51.

38. Wright J, Schult RL. Recognition and classification of nonlinear chaotic signals. Chaos. 1993 Jul;3(3):295–304.

39. Lucy LB. An iterative technique for the rectification of observed distributions. Astronomical Journal, Vol 79, p 745 (1974) [Internet]. 1974; Available from: https://adsabs.harvard.edu/full/1974AJ.....79..745L

40. Tran P, Tran A, Phung Q, Hoai M. Explore Image Deblurring via Blur Kernel Space [Internet]. arXiv [cs.CV]. 2021. Available from: http://arxiv.org/abs/2104.00317

## Appendix A: The Convolutional Architecture for VAE

The convolutional architecture for VAE uses convolutional layers for feature extraction and image reconstruction. The encoder consists of five convolutional layers and two fully connected layers; the decoder consists of one fully connected layer and five transposed convolutional layers.

*Encoder*

The encoder compresses the input image into the latent space. It consists of multiple convolutional layers, each followed by batch normalization and ReLU activation.

*Layer 1.* Reflection padding followed by a convolutional layer with 64 filters of size 3x3, stride 2, and padding 1, resulting in an output shape of [-1, 64, 128, 128]. This is followed by batch normalization and ReLU activation.

*Layer 2.* Reflection padding followed by a convolutional layer with 128 filters of size 3x3, stride 2, and padding 1, resulting in an output shape of [-1, 128, 64, 64]. This is followed by batch normalization and ReLU activation.

*Layer 3.*   Reflection padding followed by a convolutional layer with 256 filters of size 3x3, stride 2, and padding 1, resulting in an output shape of [-1, 256, 32, 32]. This is followed by batch normalization and ReLU activation.

*Layer 4.*   Reflection padding followed by a convolutional layer with 256 filters of size 3x3, stride 1, and padding 1, resulting in an output shape of [-1, 256, 32, 32]. This is followed by batch normalization and ReLU activation.

*Residual Blocks.*   Five residual blocks, each consisting of reflection padding followed by two convolutional layers with 256 filters, batch normalization, and ReLU activation. The output shape remains [-1, 256, 32, 32] throughout.

*Fully Connected Layers.*   A fully connected layer to calculate the mean of the latent distribution with an output shape of [latent_dims], and a fully connected layer to calculate the log-variance of the latent distribution with an output shape of [latent_dims].

### *Decoder*

The decoder reconstructs the image from the latent space. It uses transposed convolutions to upsample the latent representation back to the original image dimensions.

*Initial Layer.*   A fully connected layer to expand the latent space representation to match the dimensions required by the transposed convolutional layers, resulting in an output shape of [-1, 256, 4, 4].

*Layer 1.*   Transposed convolutional layer with 256 filters of size 3x3, stride 2, padding 1, and output padding 1, resulting in an output shape of [-1, 256, 8, 8]. This is followed by batch normalization and ReLU activation.

*Layer 2.*   Transposed convolutional layer with 128 filters of size 3x3, stride 2, padding 1, and output padding 1, resulting in an output shape of [-1, 128, 16, 16]. This is followed by batch normalization and ReLU activation.

*Layer 3.*   Transposed convolutional layer with 64 filters of size 3x3, stride 2, padding 1, and output padding 1, resulting in an output shape of [-1, 64, 32, 32]. This is followed by batch normalization and ReLU activation.

*Layer 4.*   Transposed convolutional layer with 1 filter of size 3x3, stride 2, padding 1, and output padding 1, resulting in the final output shape of [-1, 1, 128, 128]. This is followed by the specified decoder activation function: Tanh.

Table A. Metrics comparison.

| Metric | Input vs Target | VAE | | UNIT | |
|---|---|---|---|---|---|
| | | Output vs Target | Output vs Input | Output vs Target | Output vs Input |
| MAE (cGy) | 4.3 ± 2.4 | **4.1 ± 2.3** | 3.0 ± 1.5 | 6.4 ± 4.0 | 5.2 ± 4.0 |
| PSNR (dB) | 41.49 ± 5.17 | **42.14 ± 4.91** | 44.70 ± 4.43 | 38.89 ± 4.82 | 41.67 ± 5.32 |
| SSIM | 0.96 ± 0.02 | 0.97 ± 0.02 | 0.97 ± 0.01 | 0.97 ± 0.02 | 0.97 ± 0.02 |
| FID | 26.53 | 8.36 | 32.11 | **8.16** | 28.50 |
| **RDD (%)**: | | | | | |
|   In-field | 2.5 ± 1.2 | 2.5 ± 1.2 | 1.6 ± 0.6 | 5.5 ± 4.5 | 4.7 ± 4.6 |
|   Out-of-field | 1.8 ± 0.8 | **1.7 ± 0.7** | 1.3 ± 0.5 | 2.3 ± 1.4 | 1.9 ± 1.4 |
| **ADD (cGy)**: | | | | | |
|   In-field | 5.4 ± 2.9 | **5.3 ± 3.1** | 3.5 ± 1.7 | 10.8 ± 8.4 | 9.4 ± 8.8 |
|   Out-of-field | 4.0 ± 2.3 | **3.7 ± 2.1** | 2.9 ± 1.7 | 4.8 ± 2.8 | 3.8 ± 2.7 |

Bold - best value not taking into account the auxiliary columns (gray). Underscore - best value overall.

### Appendix B: Metrics Table

The metrics comparison table (Table A) contains several columns, each serving a specific purpose when evaluating the performance of our I2I translation approaches. The "Inputs vs Target" column demonstrates how close the predicted images are to the real measured images. This is the baseline our models have to improve. The "Output vs Target" metrics show how close the images generated by our models are to the real measured images. The "Output vs. Input" metrics show how much the generated images differ from the original predicted images. These columns are included as an auxiliary information.

All metrics, except for FID, were calculated as means of values received for individual image pairs in the test set. FID is the only metric that can be calculated for the sample set as a whole.

**Appendix C: Metrics Comparison Plots**

The provided plots illustrate metrics comparison for 100 samples. The y-axis represents units, while the x-axis corresponds to the sample index. The plots aim to visually compare the performance and accuracy of the VAE and UNIT models against the baseline (Input vs Target).

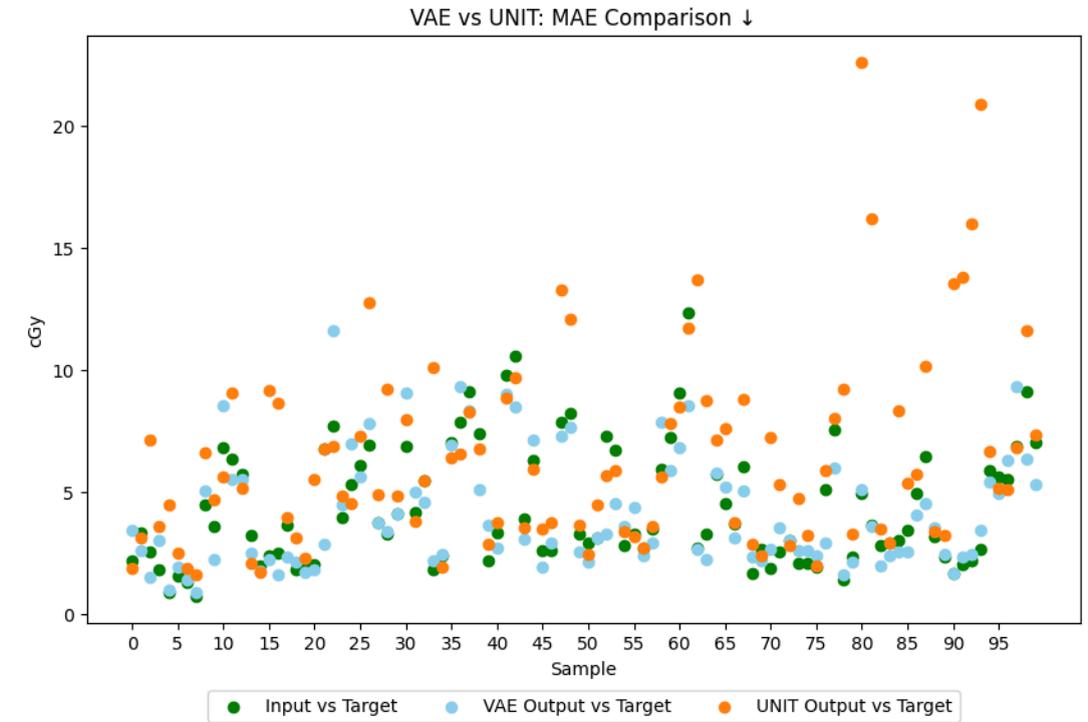

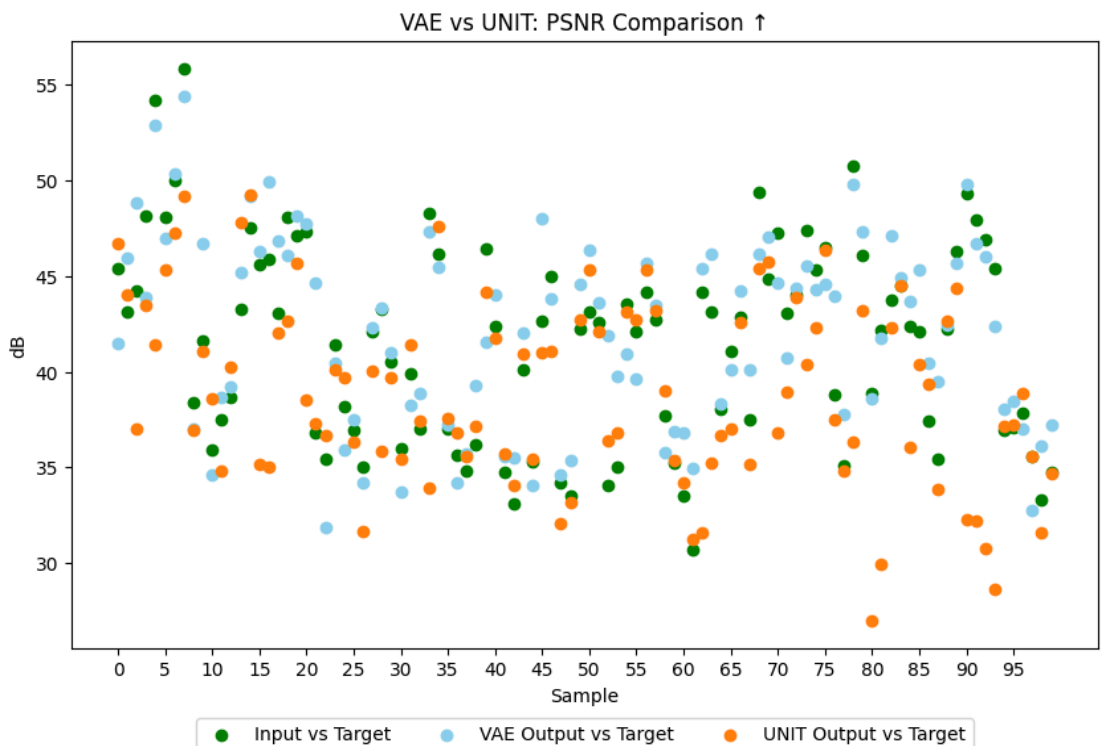

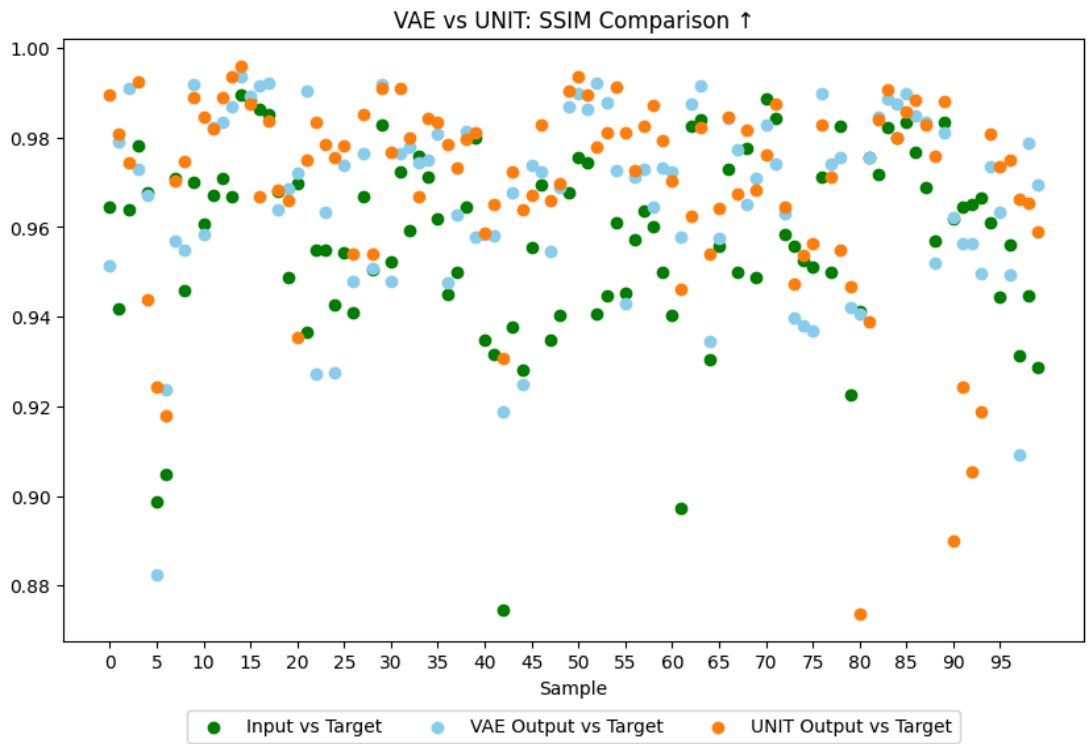

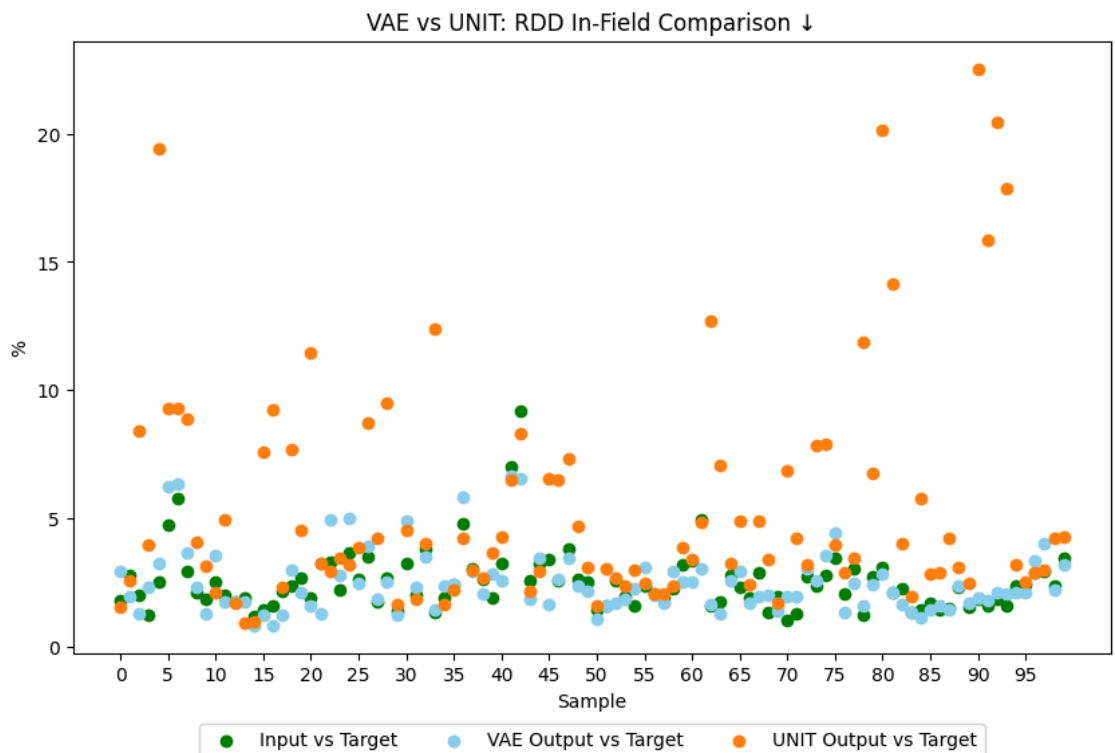

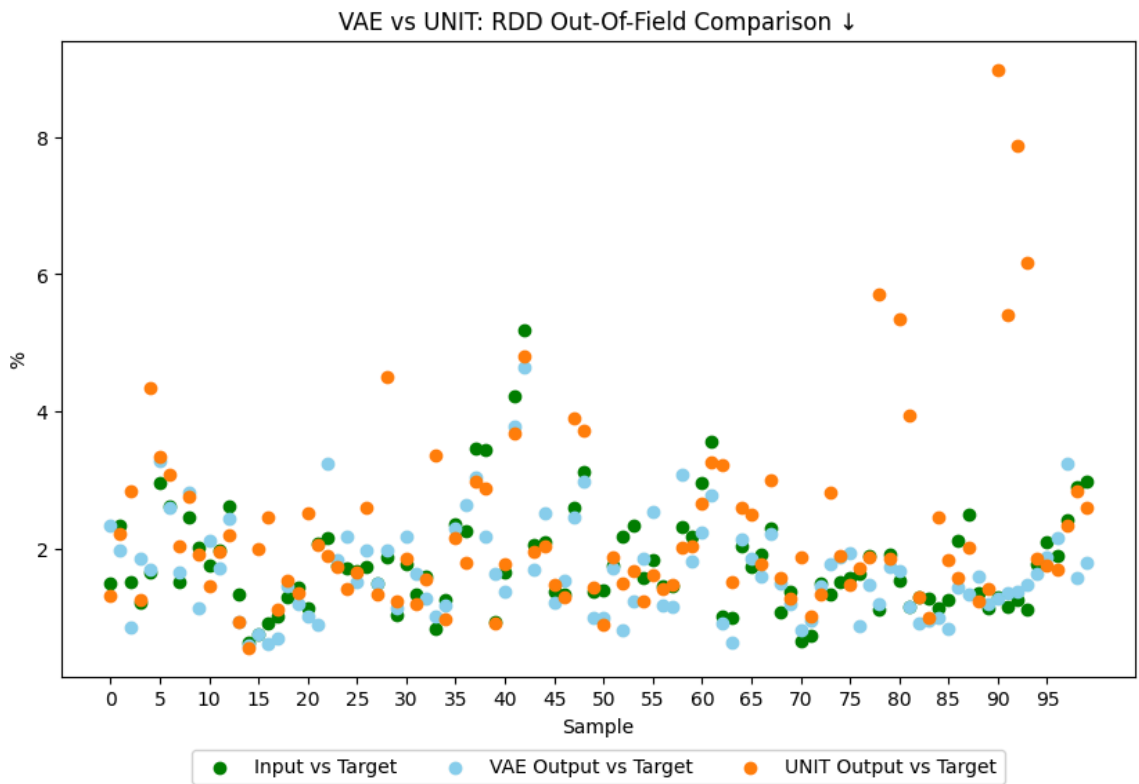

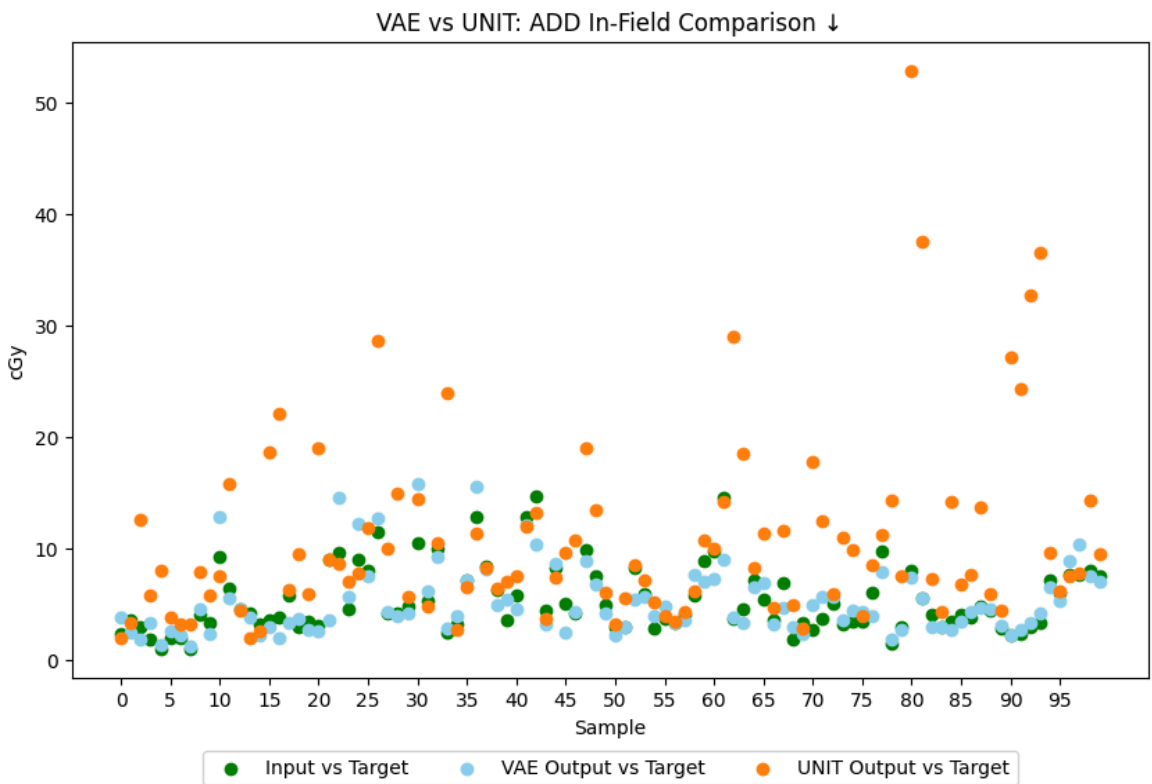

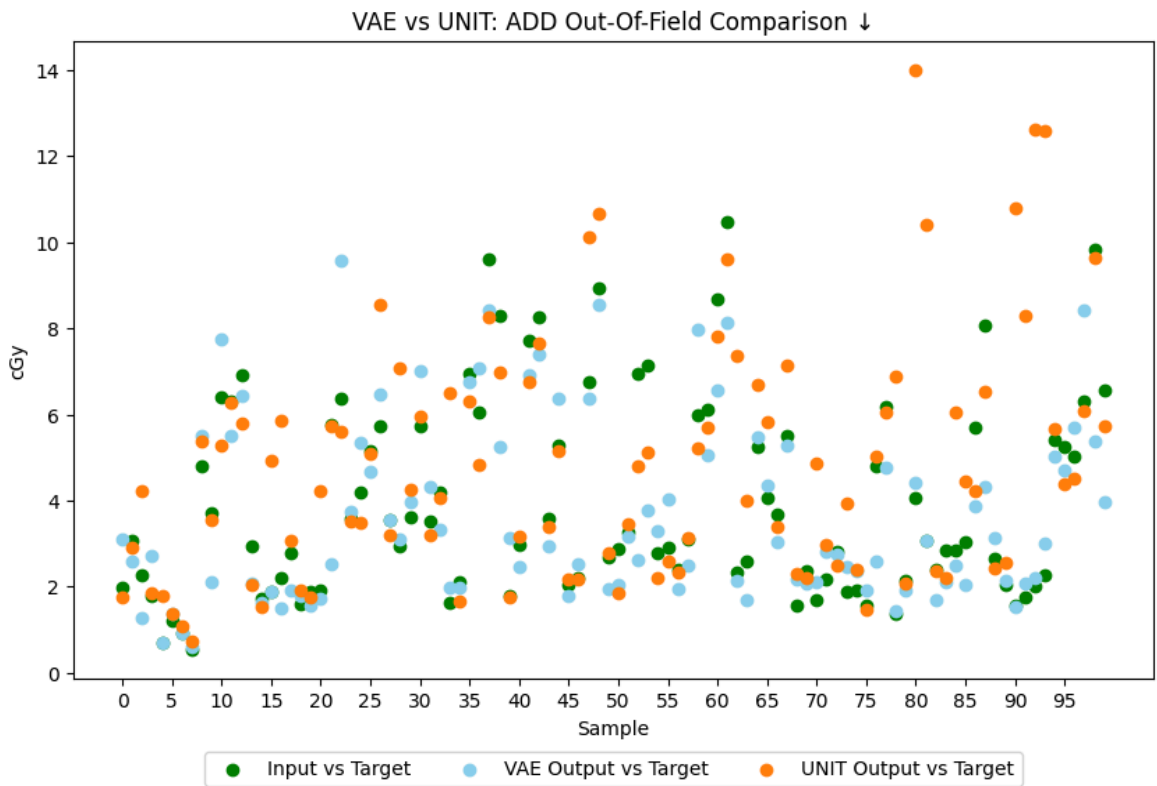